\documentclass[12pt,epsfig]{article}
\usepackage[usenames]{color}
\usepackage{epsfig}

\begin{document}

\newcommand{\be}{\begin{eqnarray}}
\newcommand{\ee}{\end{eqnarray}}
\newcommand{\inli}{\int\limits}
\newcommand{\sumli}{\sum\limits}
\newcommand{\ggam}{\gamma\gamma}
\newcommand{\cbc}{c\bar c}
\newcommand{\bbb}{b\bar b}

\newcommand{\lam}{\lambda}
\newcommand{\vk}{{\bf k}}
\newcommand{\nn}{\nonumber}

\def\la{\mathrel{\mathpalette\fun <}}
\def\ga{\mathrel{\mathpalette\fun >}}
\def\fun#1#2{\lower3.6pt\vbox{\baselineskip0pt\lineskip.9pt
\ialign{$\mathsurround=0pt#1\hfil##\hfil$\crcr#2\crcr\sim\crcr}}}

\newcommand{\bea}{\begin{eqnarray}}
\newcommand{\eea}{\end{eqnarray}}
\def\la{\mathrel{\mathpalette\fun <}}
\def\ga{\mathrel{\mathpalette\fun >}}
\def\fun#1#2{\lower3.6pt\vbox{\baselineskip0pt\lineskip.9pt
\ialign{$\mathsurround=0pt#1\hfil##\hfil$\crcr#2\crcr\sim\crcr}}}
\newcommand{\bc}{\begin{center}}
\newcommand{\ec}{\end{center}}

\title{Quasi-diffraction production of white quark--gluon clusters at
superhigh-energy hadron collisions. }

\author{
 V.V. Anisovich,
 L.G. Dakhno, M.A. Matveev,
 V.A. Nikonov, \\
\it{Petersburg Nuclear Physics Institute, Gatchina,  Russia}}

\date{\today}

\maketitle

\begin{abstract}
We discuss a collective effect, which can be possible in hadron--hadron
collisions at superhigh energies, that is, a
quasi-diffraction production of several white clusters of quarks and
gluons. Being  transformed into hadrons, such clusters are sources of
the fastest particles, which move forward nearly parallel to each
other.
 \end{abstract}
 \vspace{0.5cm}
PACS numbers: 11.55.Fv, 12.39.-x, 123.1K, 14.40.Aq

\section{Hadron collisions at high and superhigh energies }

The region of soft quark and gluon interactions can be considered
within the approach of effective particles -- constituent quarks and
massive gluons. A similar idea is used in condensed matter physics,
where effective particles and effective interactions are the operating
tools. Reviewing the experimental data \cite{WAKNS,book3}, we can see
that the hypothesis of hadrons, being composite systems of two
(meson) or three (baryon) confined constituent quarks, works well.

\begin{figure}
\centerline{\epsfig{file=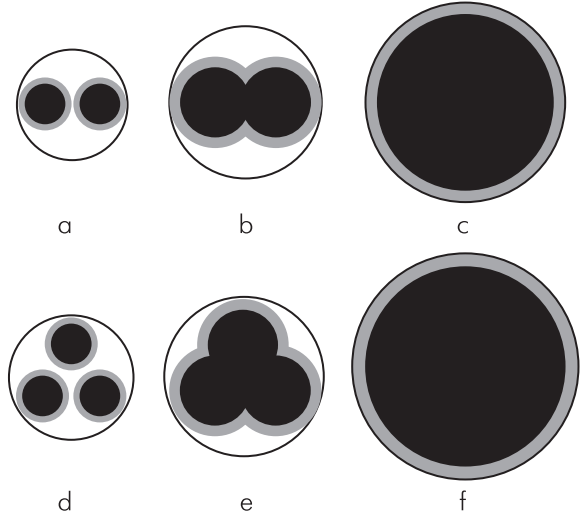,width=3in}}
\caption{Quark
structure of the low-lying meson (a,b,c) and  baryon (d,e,f) in the
constituent quark model.  At moderately high energies (a,c)
constituent quarks inside the hadron are spatially separated. With
the energy growth quarks  overlap partially (b,e); at superhigh
energies (c,f) quarks  completely overlap, and hadron collisions
lose the additivity property.  } \label{Y1f11} \end{figure}

 At low and moderately high energies the
constituent quarks inside a hadron, such as  pion or  nucleon,
are spatially separated. The constituent quark of a fast-moving
hadron is represented  as a bunch of partons. The changes, which the
clouds of colliding quarks undergo from moderately high energies up
to superhigh ones, can be demonstrated in the impact parameter space
(see Fig. \ref{Y1f11}) -- a detailed discussion is given in
\cite{WAKNS,book3}.

\begin{figure}
\centerline{\epsfig{file=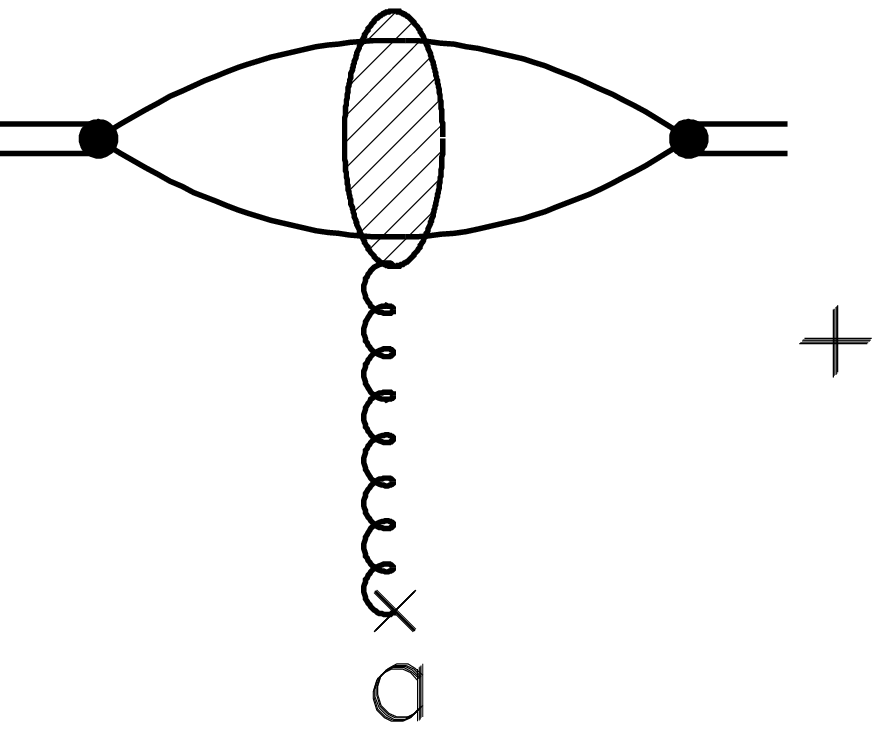,height=3.5cm}\hspace{0.5cm}
            \epsfig{file=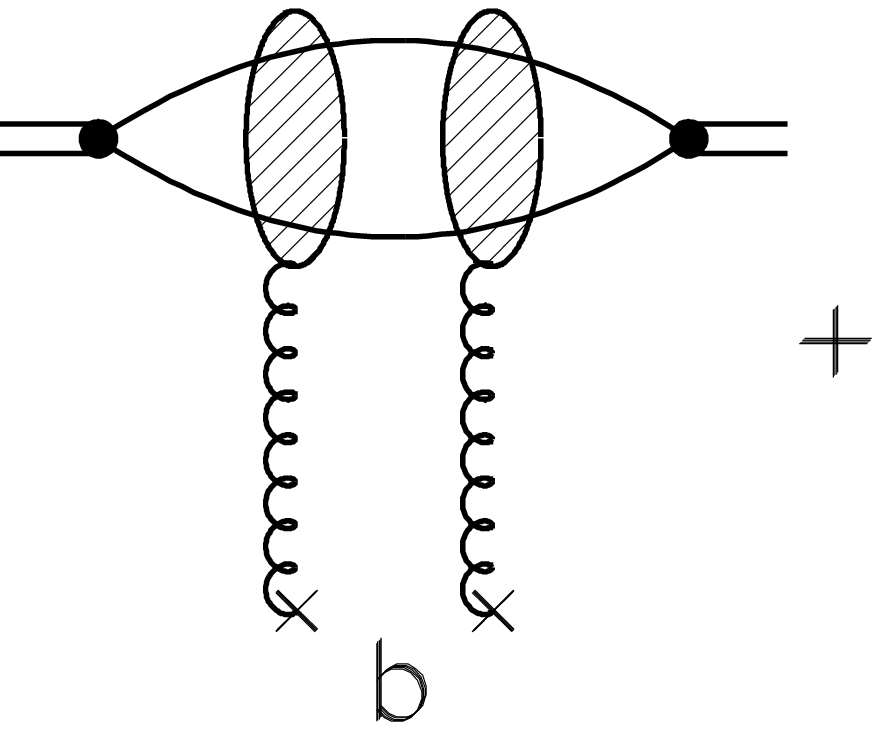,height=3.5cm}\hspace{0.5cm}
            \epsfig{file=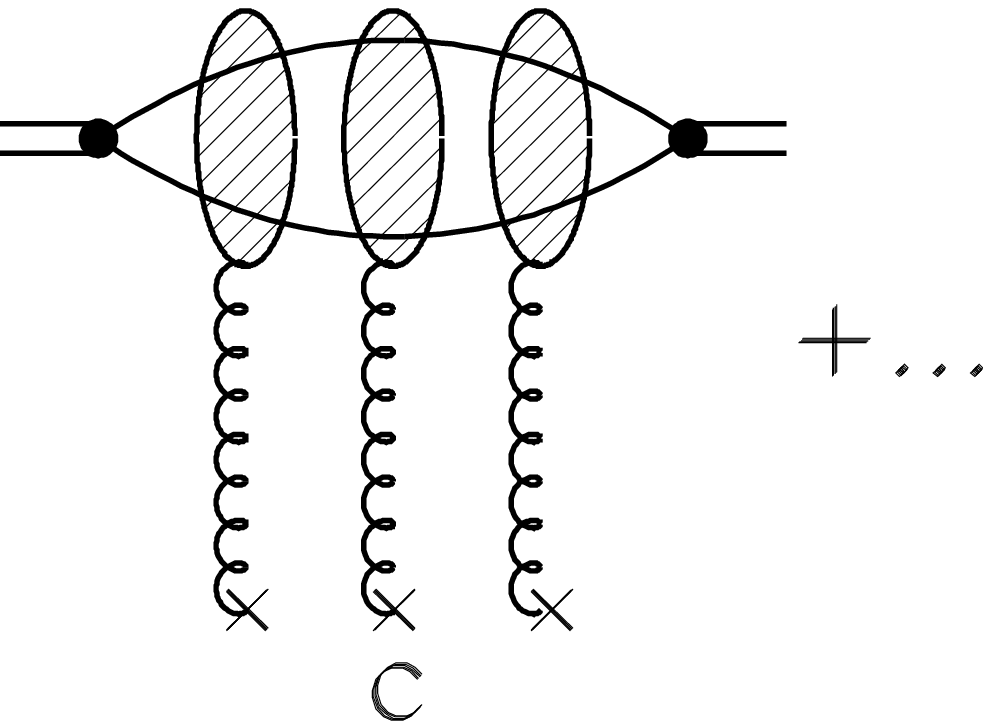,height=3.5cm}}
\caption{ Multi-pomeron exchange diagrams,
used for mesons in the eikonal approach [4].}
\end{figure}

 \begin{figure}
\centerline{\epsfig{file=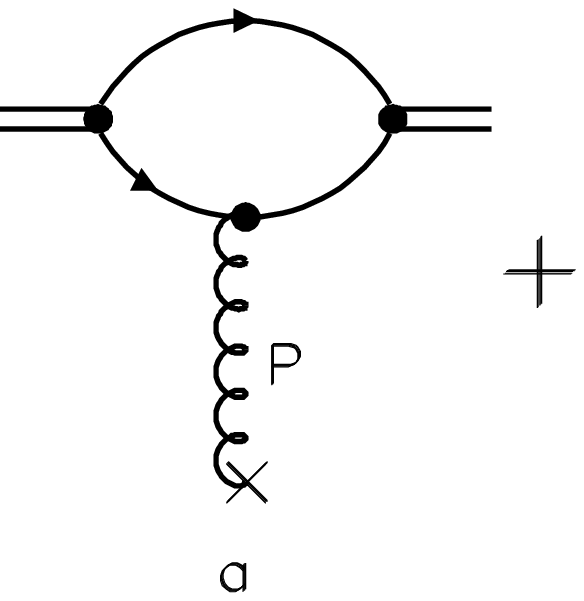,height=4.5cm}
            \epsfig{file=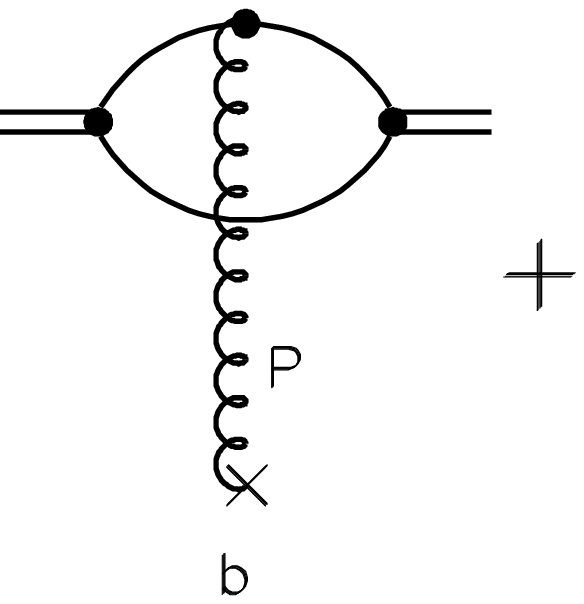,height=4.5cm}
            \epsfig{file=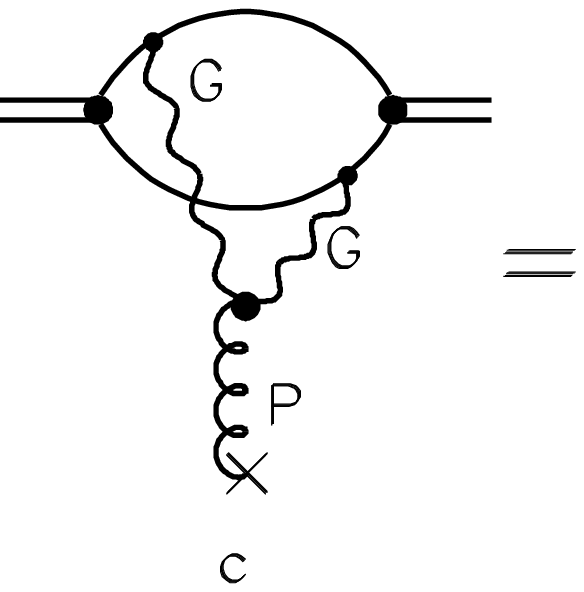,height=4.5cm}
            \epsfig{file=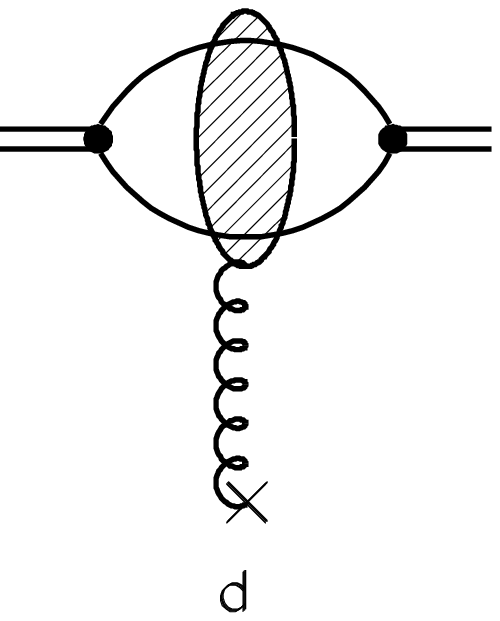,height=4.5cm}}
\caption{Vertex for pomeron--meson
interaction, with {\bf P} being a
pomeron and {\bf G} reggeized gluon used in [4]. The diagram
{\bf PGG} provides the colour screening.} \end{figure}

The figure \ref{Y1f11}a shows us a ``picture'' of a meson, while
Fig. \ref{Y1f11}d is that for a
nucleon in the impact parameter space, ({\it i.e.} it is how the
incoming hadrons look
like from the point of view of the target).
In the impact parameter
space quarks are black at moderately high energies: this follows
from the investigation of the proportion of true inelastic and
quasi-inelastic processes \cite{YALR}. Accordingly, in Figs.
\ref{Y1f11}a and
\ref{Y1f11}d two (for a meson) and three (for a baryon) black disks are
drawn. But the transverse  disk size
increases, and at intermediate energies ($p_{lab} \sim 500-1000$
GeV/c) the quarks partially overlap (Figs. \ref{Y1f11}b, \ref{Y1f11}e).
In this
energy region the additivity may already be broken in the collision
processes. Furthermore, there is a complete overlapping of
clouds (Figs. \ref{Y1f11}c, \ref{Y1f11}f) and, in principle, the
meson--proton and proton--proton cross sections grow up similarly,
$\sigma_{tot}(p\bar{p})/\sigma_{tot}(\pi p)\to 1$ at $s\to \infty$.


The estimate of  characteristics of meson--nucleon and
nucleon--nucleon cross sections at increasing energies was carried
out in \cite{YDN} within the framework of eikonal approximation,
accounting  for the $t$-channel supercritical pomeron ($\bf P$ with
$\alpha_{\bf P}(0)\sim 1.29)$ and
gluon--gluon--pomeron exchanges
($\bf{GGP}$  with  $\alpha_{\bf G}(0)\sim 1.0$).

In Figs. 2,3 we demonstrate eikonal interaction for meson, analogous
interaction is written for the nucleon.
 For energies 0.5 TeV$\leq \sqrt{s}\leq$ 20 TeV the
cross sections calculated in this way  behave as follows \cite{YDN}:
\begin{eqnarray}
\label{Y7.1} & &
\sigma_{tot}(p\bar{p})=\bigg (
49.80+8.16\ln\frac{s}{9s_0}+0.32\ln^2\frac{s}{9s_0}\bigg ) \, {\rm mb}
\ , \qquad s_0=10^4\,  {\rm GeV}^2 \ ,     \nonumber
\\ & & \sigma_{tot}(\pi
p)=\bigg (
30.31+5.70\ln\frac{s}{6s_0}+0.32\ln^2\frac{s}{6s_0}\bigg ) \, {\rm mb}  \,
,
\end{eqnarray}
and
\begin{eqnarray}
\label{Y7.1a}
& &
\sigma_{el}(p\bar{p})=\bigg (
8.19+3.027\ln\frac{s}{9s_0}+0.16\ln^2\frac{s}{9s_0}\bigg ) \, {\rm mb}
\ ,      \nonumber
\\ & &
\sigma_{el}(\pi p)=\bigg (
3.87+1.567\ln\frac{s}{6s_0}+0.16\ln^2\frac{s}{6s_0}\bigg ) \, {\rm mb}
\, .
\end{eqnarray}

\begin{figure}[ht]
\centerline{\epsfig{file=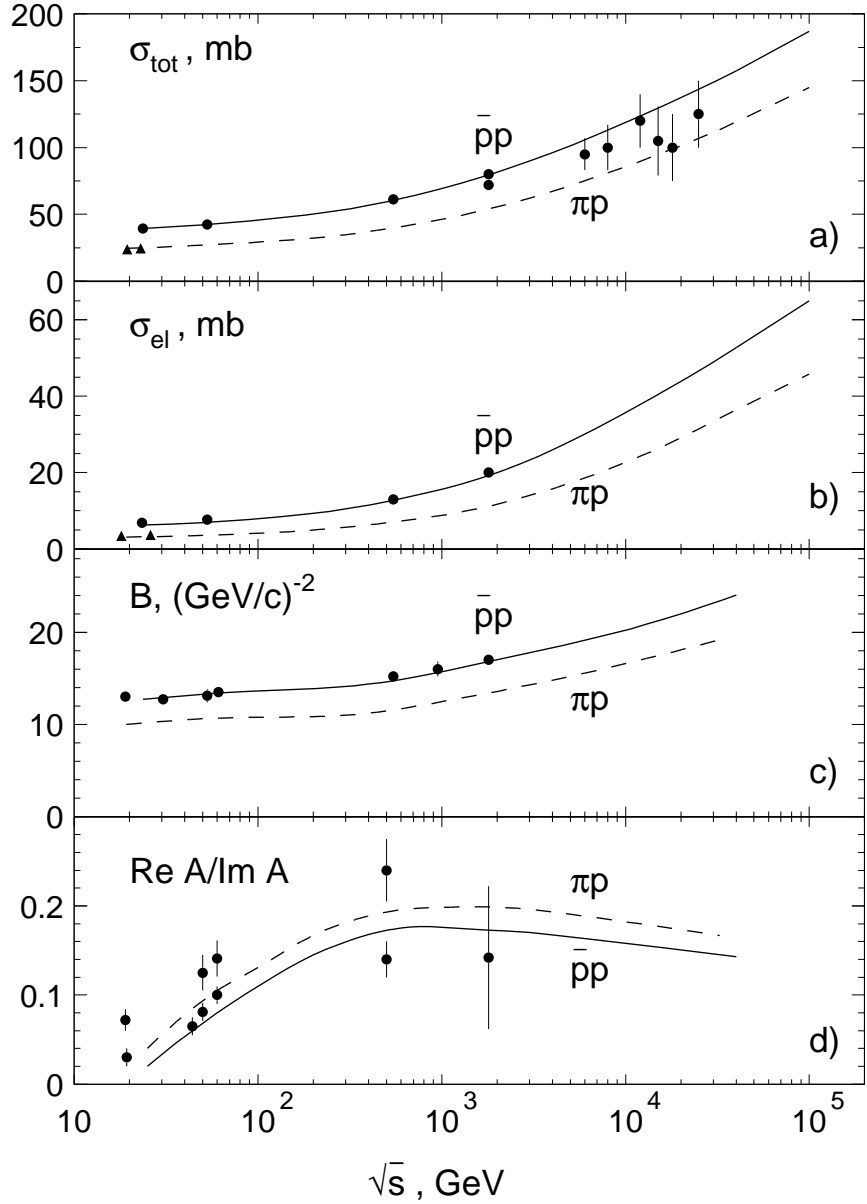,width=4.8in}}
\caption{Calculations [4]: a) total cross sections  for $p\bar p$ and
$\pi p$ collisions, b) elastic
 cross sections, c) diffraction scattering slopes
 $d\sigma_{el}/dq^2_\perp =B\sigma_{el}\exp[-Bq^2_\perp] $,  and d)
ratios of real/imaginary parts of the scattering amplitudes at
 $q^2_\perp=0$.
 \label{sss2} }
\end{figure}

The calculated cross sections and other characteristics are shown
in Fig. \ref{sss2}. In the region
$\sqrt{s}\sim 10^4$ GeV the cosmic ray data are demonstrated, see
\cite{cosmic} and references therein.
In the region of LHC energies
($\sqrt{s}=$16 TeV), calculations \cite{YDN} predict:
\begin{equation} \label{Y7.2}
\sigma_{tot}(p\bar{p})=
131\; \mbox{mb}, \quad \sigma_{el}(p\bar{p}) = 41\; \mbox{mb} \, .
\end{equation}
As is seen, at LHC energies the asymptotic value
$\sigma_{tot} \simeq  2\sigma_{el} \,$
is not reached yet.

At LHC
energies another consequence of the quark overlap may reveal itself.
We mean the scaling violation  of proton and meson spectra in the
fragmentation region, at $p/p_{max}=x\sim 1/2 - 2/3$.
The hadron spectra have to decrease in this
region due to the loss of quark additivity in the  collision process.

\clearpage

\section{Quasi-diffraction production of white quark--gluon\\ clusters
at superhigh energies }

\begin{figure}
{\centerline{\epsfig{file=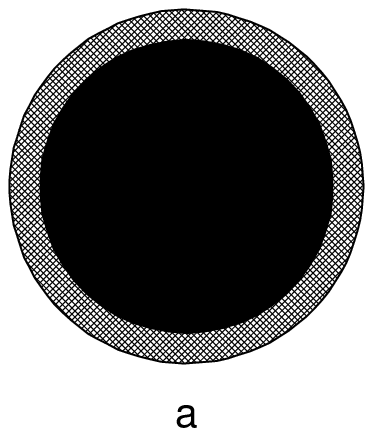,width=4cm}
           {\epsfig{file=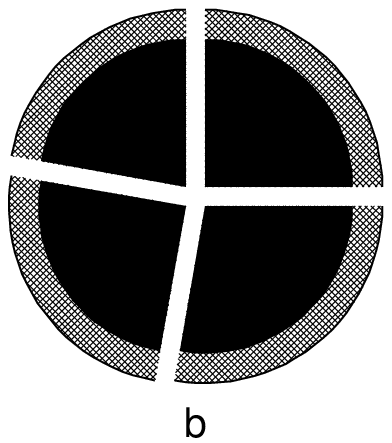,width=4cm}} }}
\caption {Variants of the behaviour of the parton black disk at
superhigh energies: (a) the hadron (black disk) grows with simultaneous
increase of the confinement radius, (b) the black disk increases
 but the growth of the confinement radius
is slower, and partons split softly into several white domains
(conventionally, we have separated white domains of the disk
by white strips). \label{Q9fig1}}
\end{figure}

 We know that the cross section $\sigma_{tot}(p\bar p)$
grows with   energy, and we are almost sure that it
will continue to grow up as $\ln^2s$.
 How does it affect the
confinement phenomenon? Does the confinement radius also
increase allowing the hadron to be a black disk (Fig.
\ref{Q9fig1}a) or the parton disk breaks
off into a number of white domains (Fig. \ref{Q9fig1}b)?

\begin{figure}
{\centerline{\epsfig{file=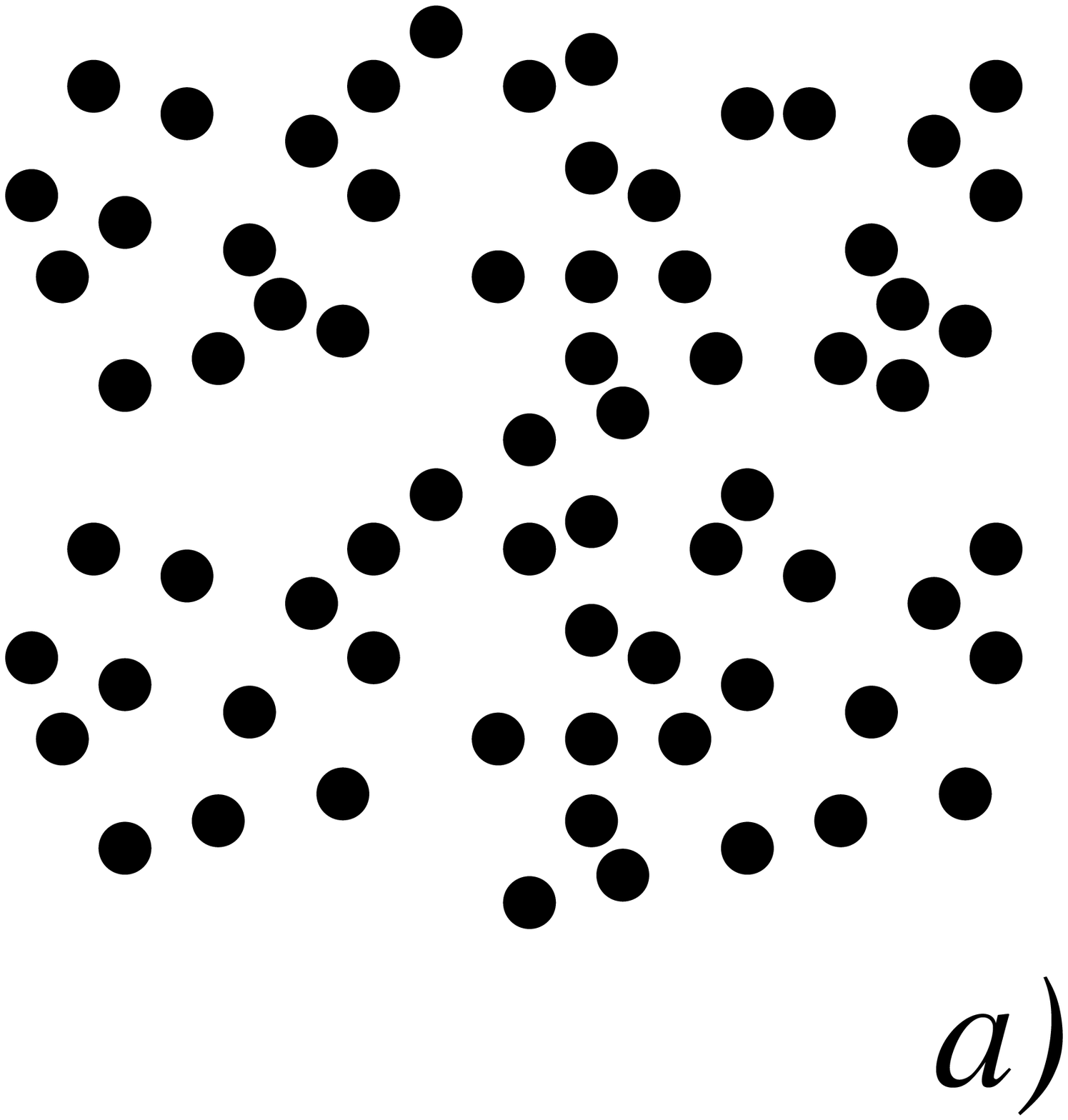,width=4cm}\hspace{0.8 cm}
             \epsfig{file=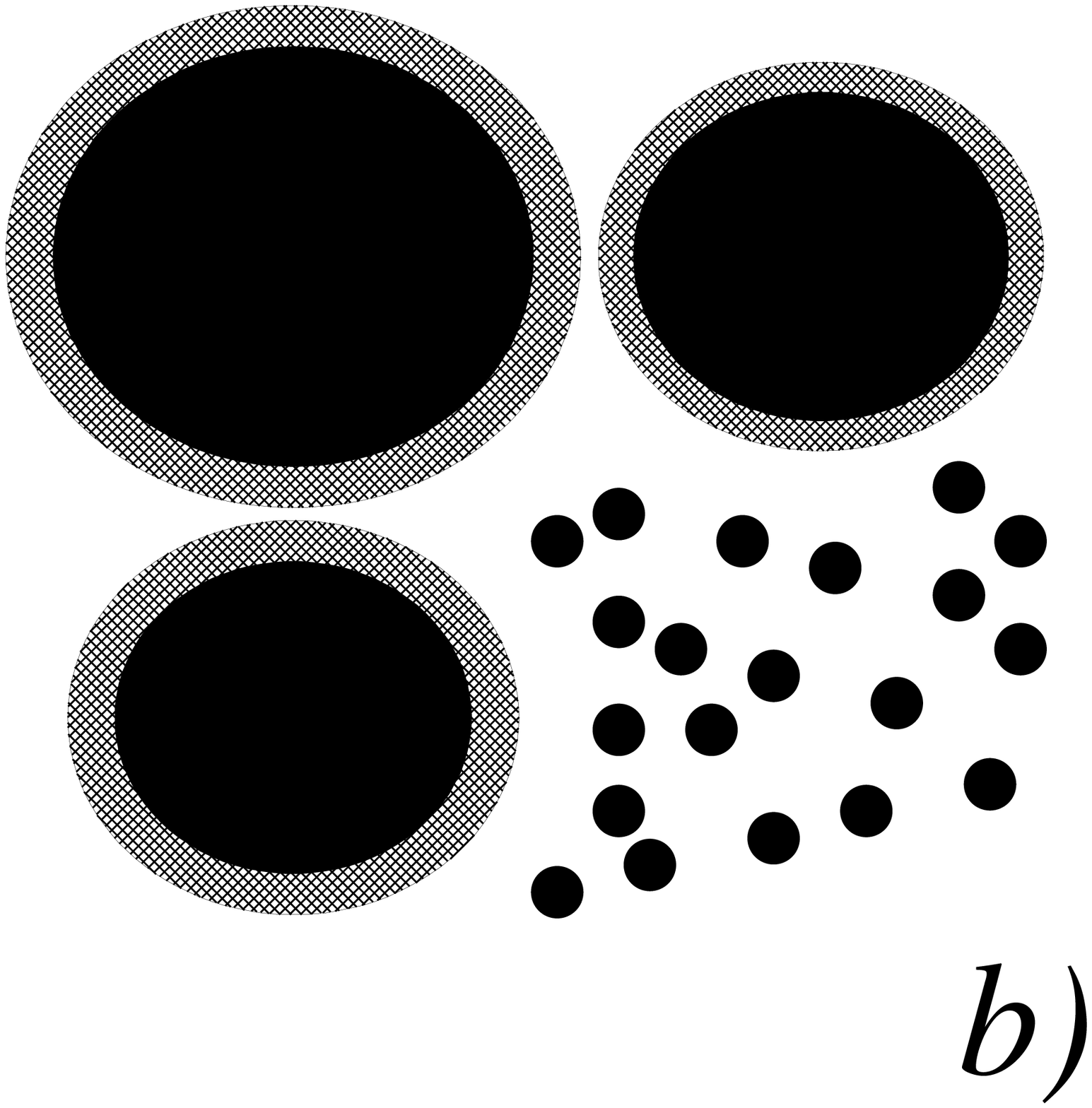,width=4cm}\hspace{0.8 cm}
             \epsfig{file=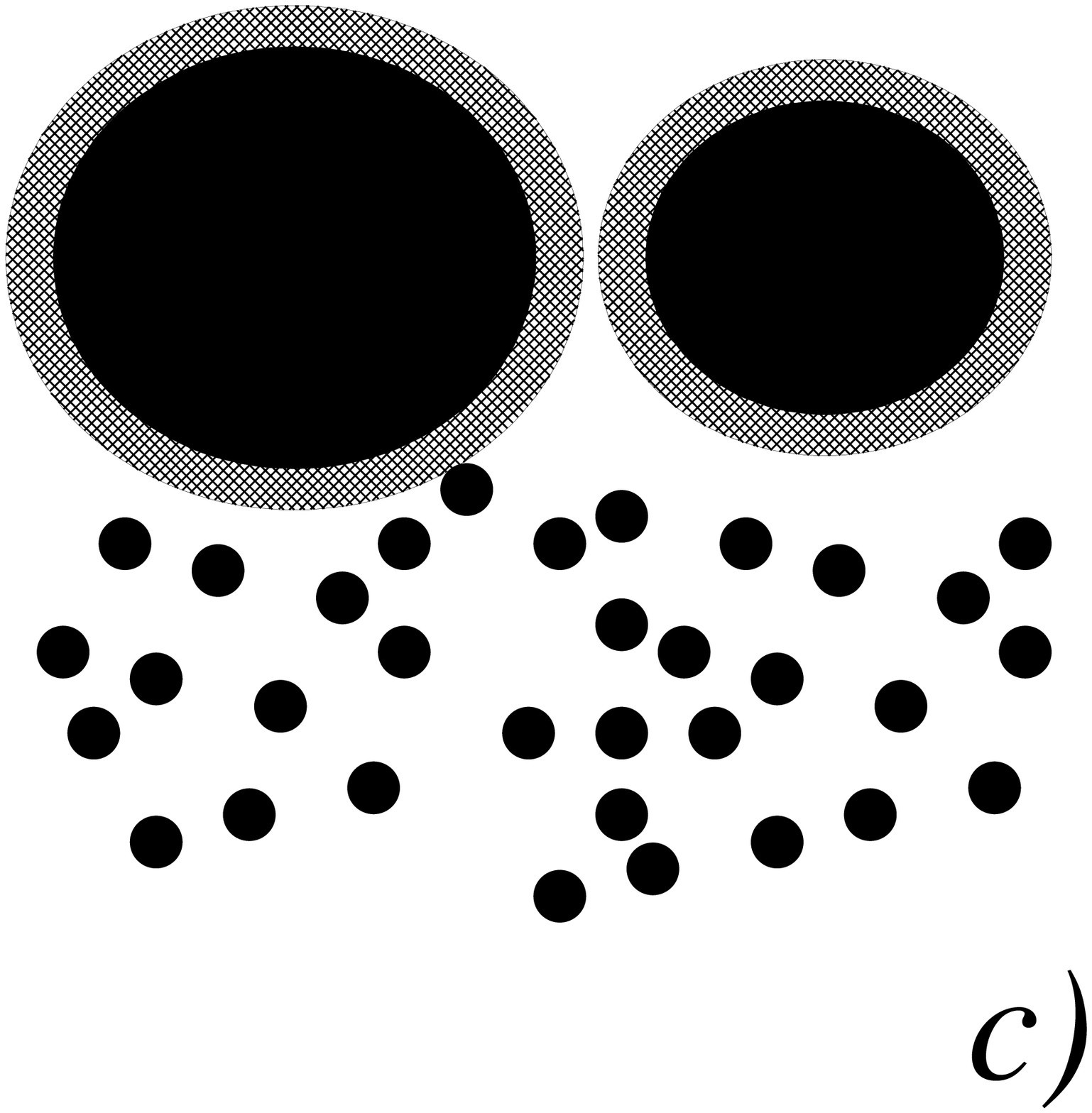,width=4cm} } }
\caption {Probable parton distributions after collision of the disk
5a with a target (6a), and disk 5b with a target (6b,6c); the pictures
6b and 6c describe  quasi-diffraction production of
several white clusters of quarks and gluons. Clusters are
moving forward being nearly parallel and transforming softly into
hadrons.
 \label{Q9fig2}}
\end{figure}

The most probable result of the interaction of the disk Fig.
\ref{Q9fig1}a with target is shown in Fig. \ref{Q9fig2}a: it is a
complete dissociation of the comb of partons with a subsequent
transformation of them into the comb of hadrons.

If at superhigh energies the standard picture is such as shown in Fig.
\ref{Q9fig1}b, the interaction can destroy not every white
domains of partons, and  a large probability
 we face  pictures of Figs. \ref{Q9fig2}b,c
type. In these
variants of dissociation, the objects to study are white clusters  of
partons. They are flying farther in $p_z$-direction separately and
almost parallel, being not "disturbed" by the soft collision with target.

To be more illustrative,
let us re-draw the pictures of Figs. \ref{Q9fig1}a,b and
Figs. \ref{Q9fig2}a,b,c by using diagrammatic language.

We believe that the physics of
picture 5a is not enigmatic: this is a comb of partons with the average
multiplicity $<n_{parton}> \sim 10^2-10^3$. If the coherence is
not broken, the parton comb turns into the proton again. Soft interaction
with the target violates the parton coherence, thus leading to the
creation of the hadron comb, with the same order of multiplicity
$<n_{hadron}> \sim 10^2-10^3$.

The process shown in Fig. 5b provides us
with a  diversity of hadron production schemes. First, let us note
that the white parton domain structure can result in the process similar
to that given in Fig. 5a -- to the production of a common comb with
the multiplicity $<n>_{parton} \sim 10^2-10^3$.

But, as was said above,
with a noticeable probability partons shown in  Fig. 5b may evolve
in  such a way that white domains do not lose their
coherence --  the self-energy part shown in Fig. 7a demonstrate just
this case; corresondingly, on the diagrammatic language, we have four
combs of partons, Fig. 8a.

Diagrams with soft interaction shown in Figs. 7b,c
give rise to processes of Figs. 8b,c (or of Fig. 6b,c). Here one
or two virtual hadrons have lost the coherent structure of their
partons and dissipated into a number of hadron states.

\begin{figure}
{\centerline{\epsfig{file=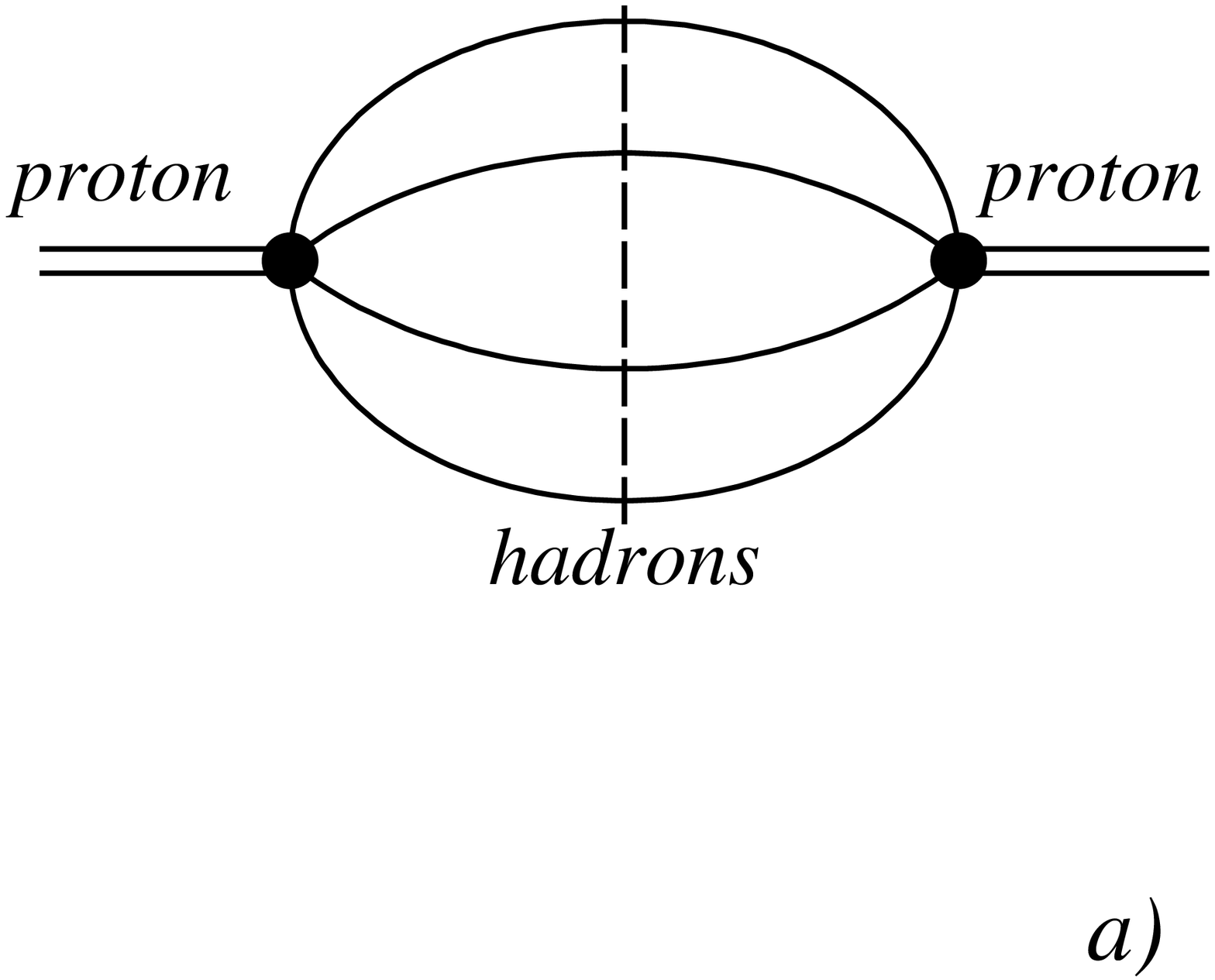,width=5cm}
             \epsfig{file=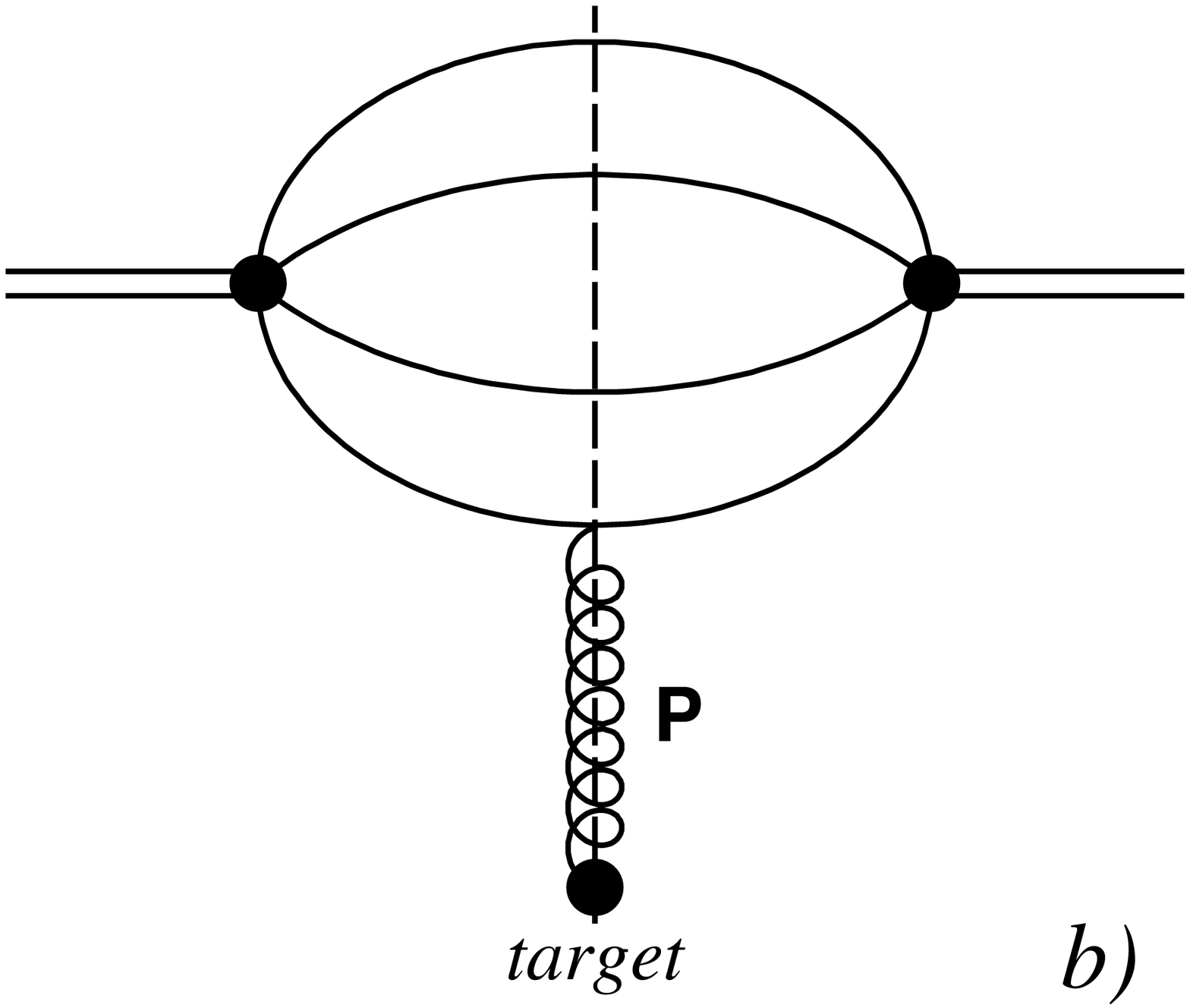,width=5cm}
             \epsfig{file=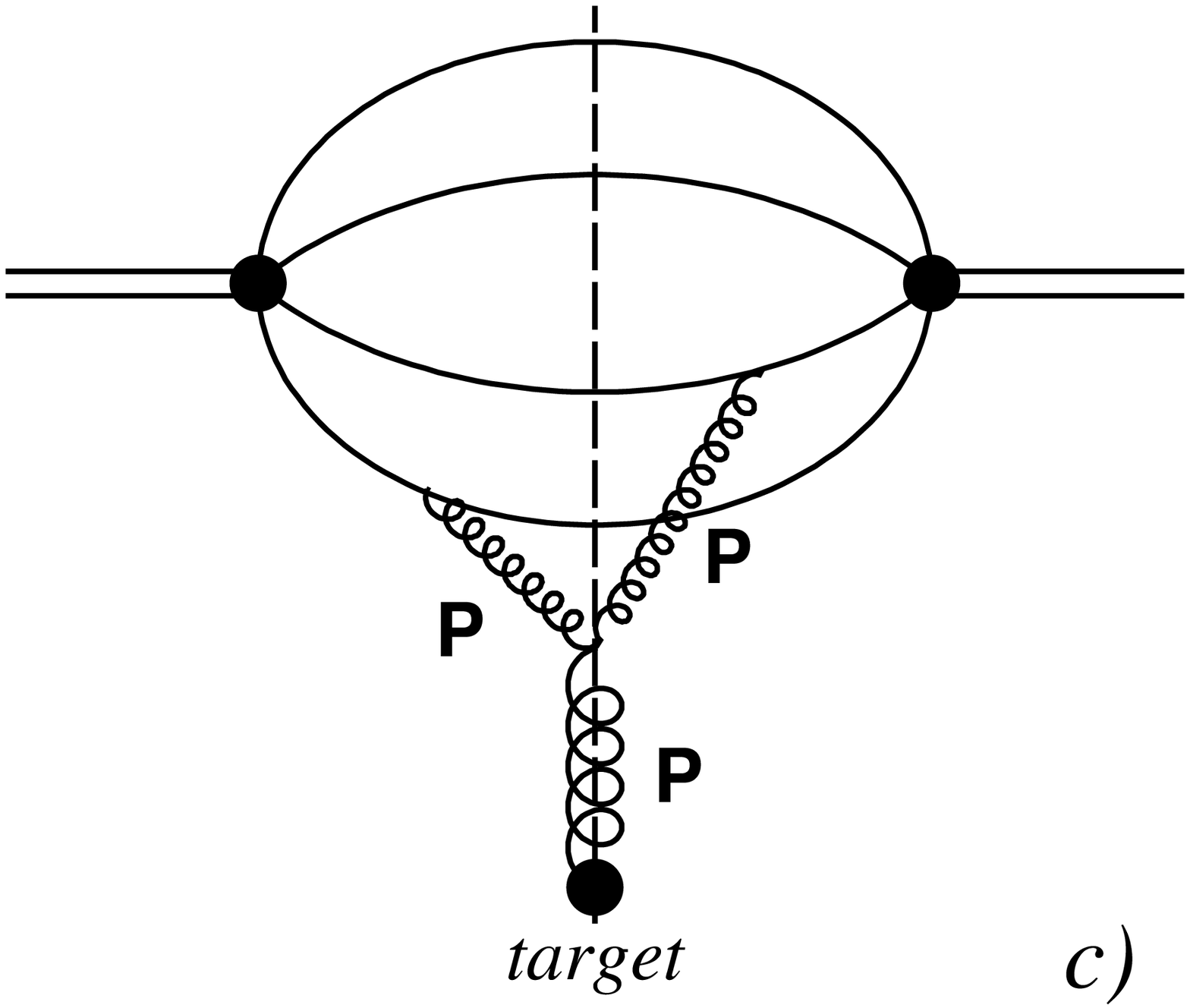,width=5cm}} }
\caption {The cutting of diagrams (dotted lines) which
illustrate  Figs. 5 and 6. a) Self-energy block of the
fast moving proton, which corresponds to four white domains in Fig.
\ref{Q9fig1}b. b) Pomeron (${\bf P}$) interaction  with one hadron from
the self-energy block, other
three hadrons (or white parton domains in Fig. 6b) do not interact.
c) Three-pomeron (${\bf PPP}$) interaction
with two hadrons from the self-energy block;  two other white parton
domains (Fig.6c) are transformed into hadrons.
\label{7} } \end{figure}

\begin{figure}[hbt]
{\centerline{\epsfig{file=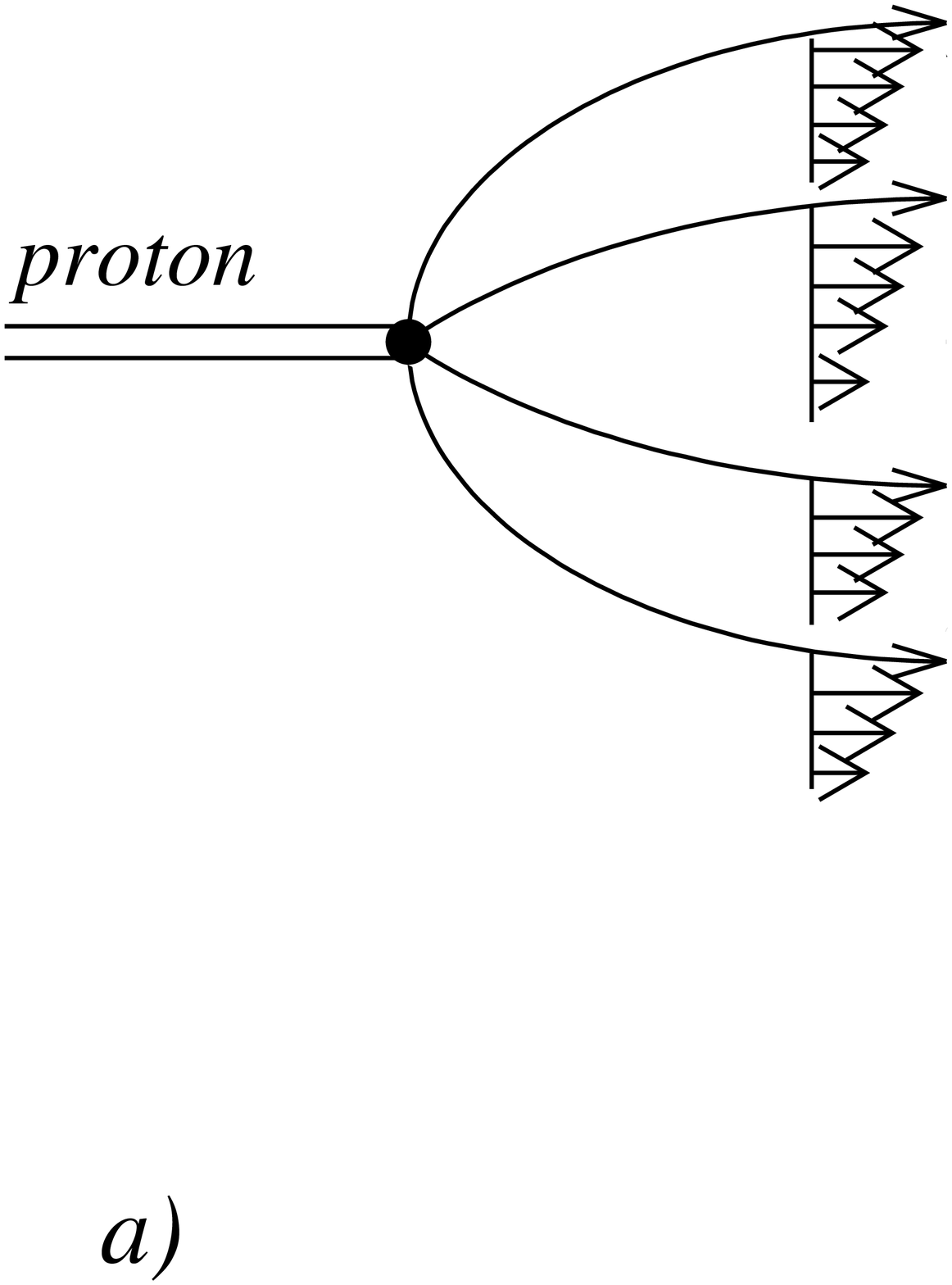,width=4cm}
             \epsfig{file=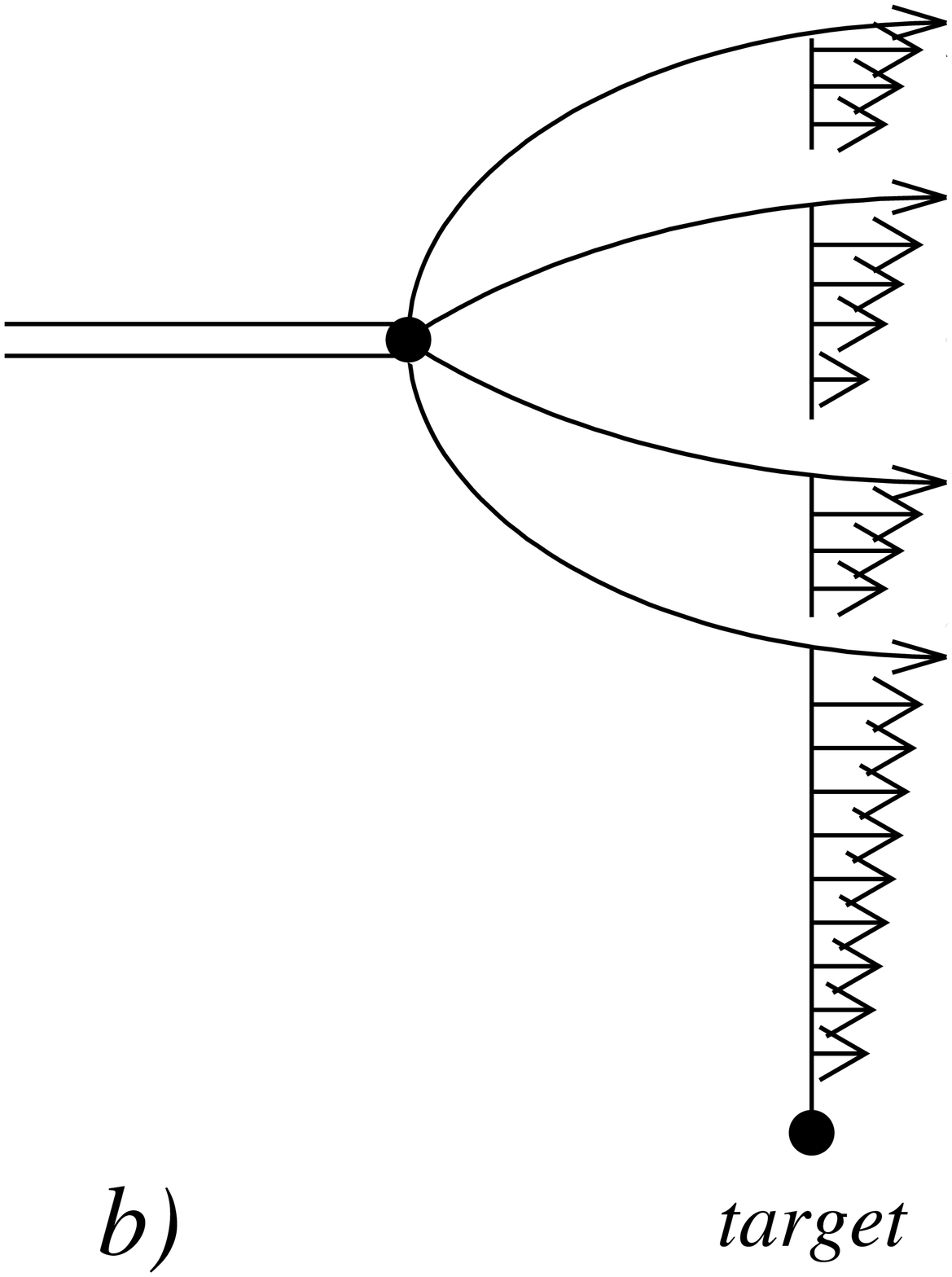,width=4cm}
             \epsfig{file=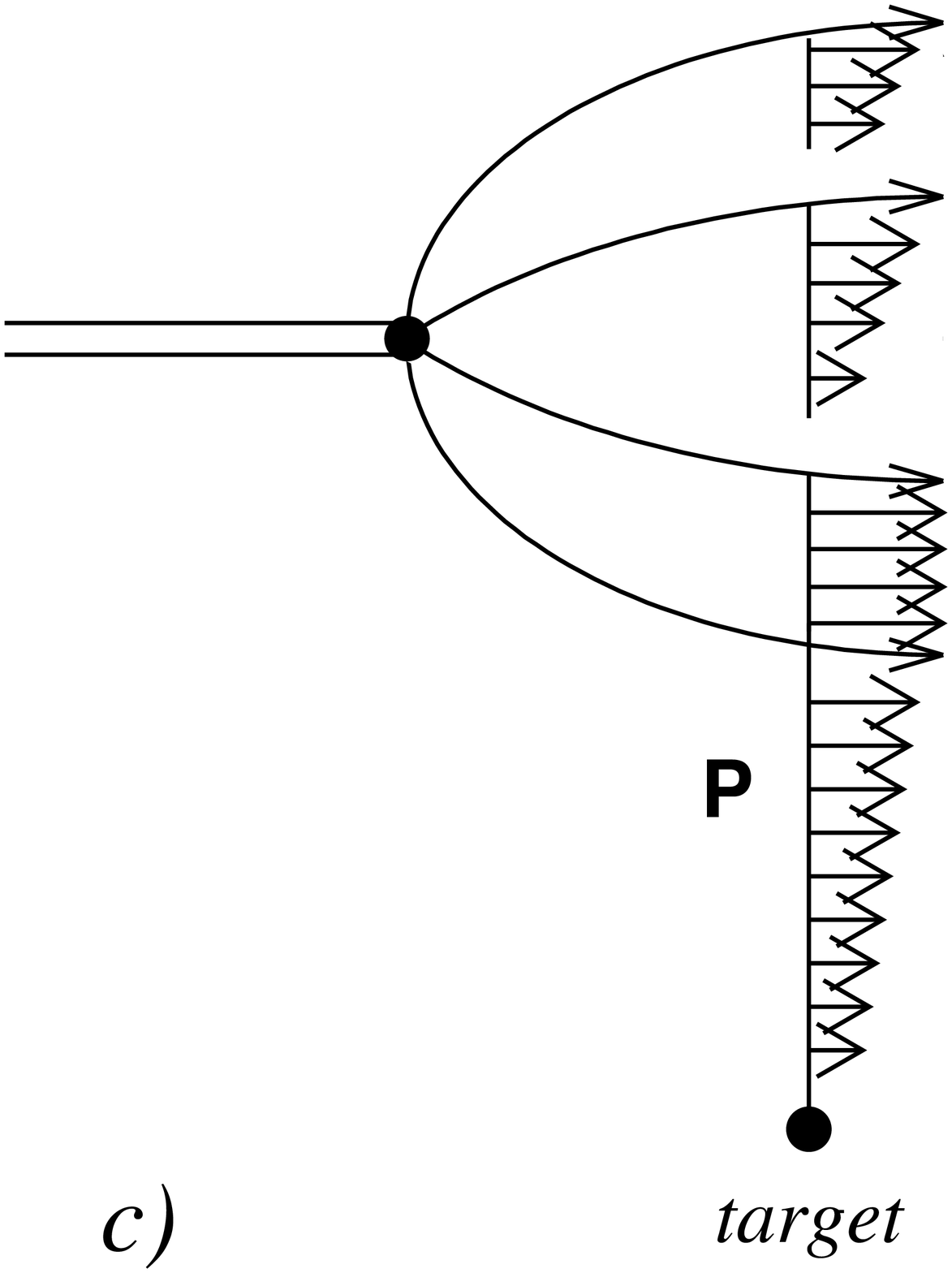,width=4cm}} }
\caption {Left-hand side blocks of diagrams of Fig.7 after the cutting
(dotted lines):
a) Four combs of partons (four white clusters of Fig.
5b) originated after cutting  the diagram 5a.
 b) Three combs of partons (three white clusters of Fig. 6b) and a
comb of pomeron partons  originated after cutting  the diagram 7b.
c) Partons originated after cutting the diagram 7c: two combs,
corresponding to white domains of Fig. 6c, and that  related to
{\bf PPP}-block. \label{8}} \end{figure}

Of course, the discussed correlated domains may be either produced or
not -- such a problem requires further investigations. In the recently
performed experiment at LHC \cite{5}, a sort of pair correlations of
hadrons have been observed, which, or at least a part of them, could
have such a domain origin. To reveal the domain origin of produced
hadrons, one should carry out more detailed study.

Namely, it is necessary to measure the $z$ distribution of particles
moving in forward direction (that is particles with small $p_\perp$).
The events with noticeable gaps as shown in Fig. 9b,c
should indicate the existence of coherent domain of partons, of the
type of shown in Figs. 6b,6c. One can expect that a number of
events with high-$p_z$ gaps must increase with energy, if the scheme
of Fig. 5b is realized in hadron collisions in the TeV region.


\begin{figure}[hbt]
\centerline{\epsfig{file=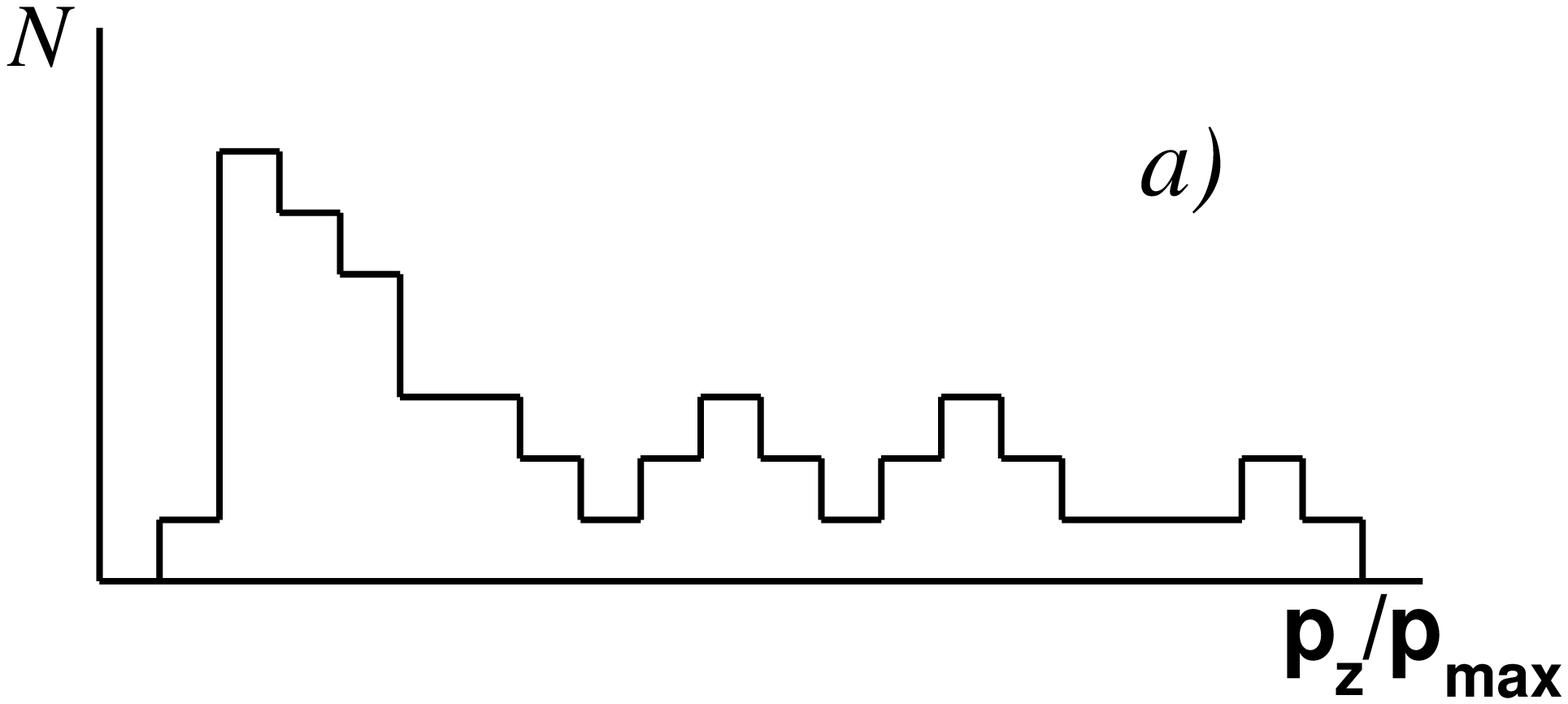,width=7cm}
            \epsfig{file=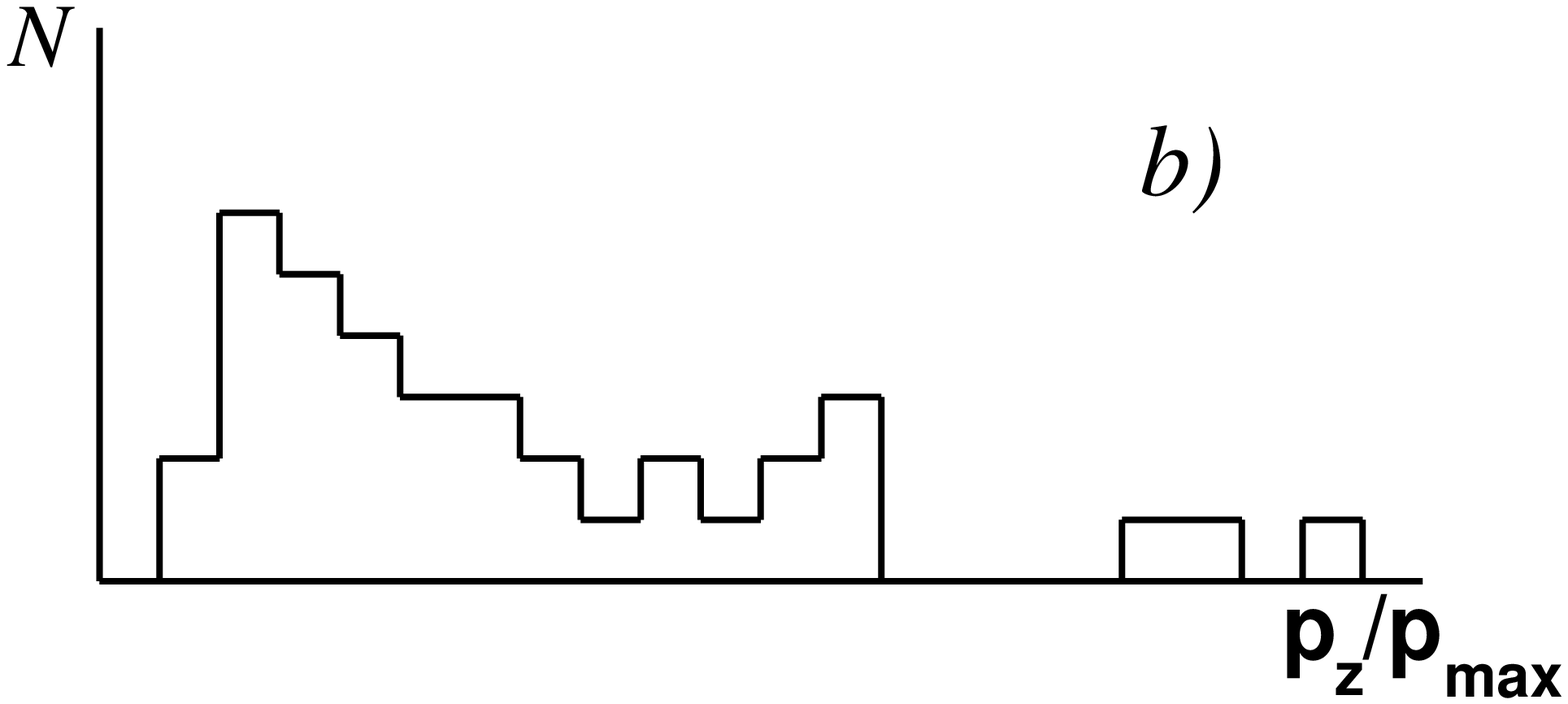,width=7cm}}
\centerline{\epsfig{file=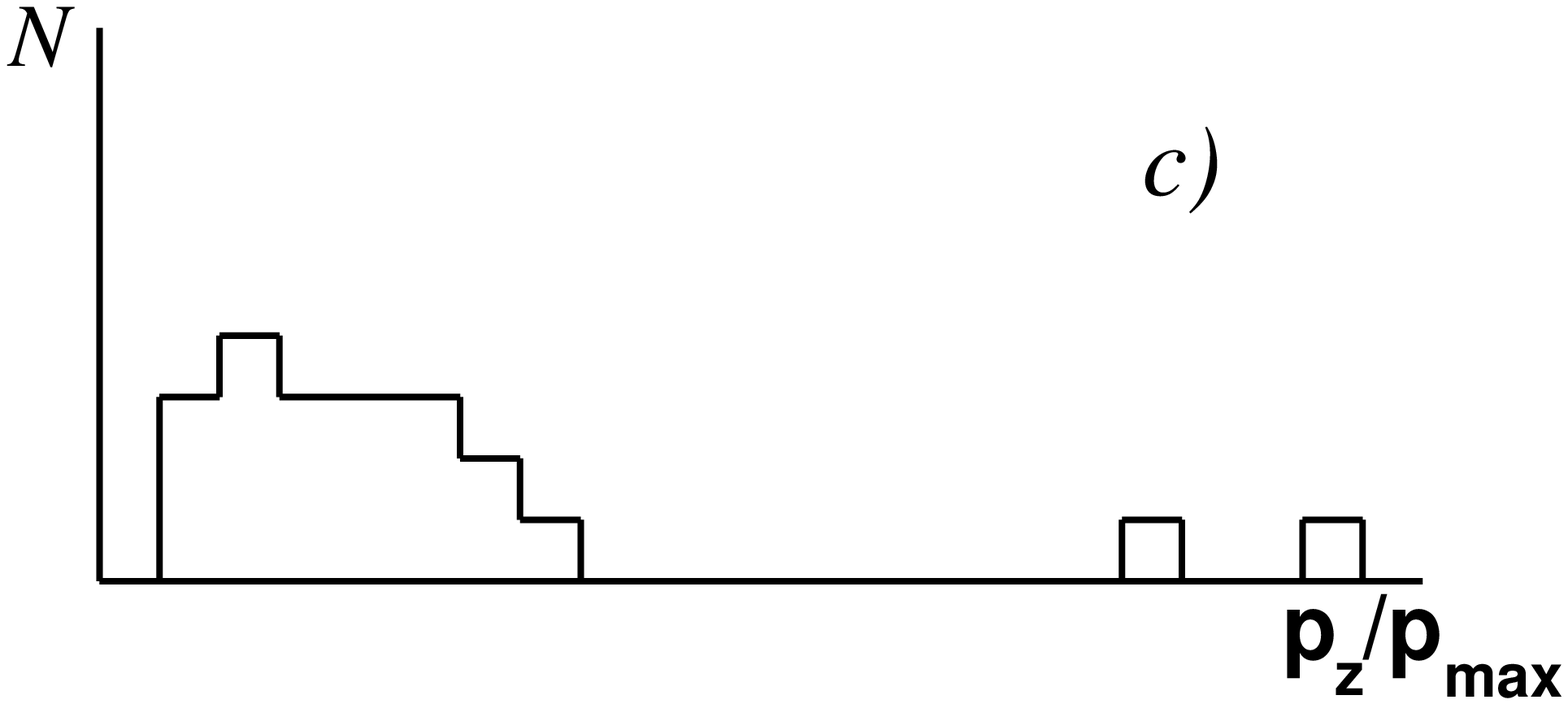,width=7cm}}
\caption {Possible $p_z$ distributions of secondaries in
processes, which correspond to Fig. 6b,c (or to cuttings of diagrams in
Figs. 8b,c). Typical property of distributions b) and c) is that hadrons
with a large fraction of $x=p_z/p_{initial}$ are separated from ofher
secondaries. } \end{figure}

\section*{Acknowledgement}

The authors are indebted to A.V. Anisovich,
Y.I. Azimov
J. Nyiri, M.G. Ryskin, A.V. Sarantsev for usefull discussions. The
paper is partly supported by grant RSGSS-3628.2008.2.

 \end{document}